# Pen-and-paper Rituals in Service Interaction: Combining High-touch and High-tech in Financial Advisory Encounters

MATEUSZ DOLATA, University of Zurich
DORIS AGOTAI, University of Applied Sciences and Arts Northwestern Switzerland
SIMON SCHUBIGER, University of Applied Sciences and Arts Northwestern Switzerland
GERHARD SCHWABE, University of Zurich

Advisory services are ritualized encounters between an expert and an advisee. Empathetic, high-touch relationship between those two parties was identified as the key aspect of a successful advisory encounter. To facilitate the high-touch interaction, advisors established rituals which stress the unique, individual character of each client and each single encounter. Simultaneously, organizations like banks or insurances rolled out tools and technologies for use in advisory services to offer a uniform experience and consistent quality across branches and advisors. As a consequence, advisors were caught between the high-touch and high-tech aspects of an advisory service. This manuscript presents a system that accommodates for high-touch rituals and practices and combines them with high-tech collaboration. The proposed solution augments pen-and-paper practices with digital content and affords new material performances coherent with the existing rituals. The evaluation in realistic mortgage advisory services unveils the potential of mixed reality approaches for application in professional, institutional settings. The blow-by-blow analysis of the conversations reveals how an advisory service can become equally high-tech and high-touch thanks to a careful ritual-oriented system design. As a consequence, this paper presents a solution to the tension between the high-touch and high-tech tendencies in advisory services.

CCS Concepts: • **Human-centered computing** → **Collaborative interaction;** *Mixed / augmented reality*; Computer supported cooperative work; **Social and professional topics** → **Socio-technical systems**

**KEYWORDS**

Advisory Service Scenario; LivePaper; Pen-and-Paper Interfaces; Interaction Rituals

**ACM Reference format:**

Mateusz Dolata, Doris Agotai, Simon Schubiger, and Gerhard Schwabe, 2019. Pen-and-paper rituals in service interaction: Combining high-touch and high-tech in financial advisory encounters. In *Proceedings of the ACM on Human-Computer Interaction*, Vol. 3, No. CSCW, Article 224, November 2019. *ACM, New York, NY, USA, 22 pages.*

https://doi.org/10.1145/3359326

This work is supported by the *Innosuisse − Swiss Innovation Agency*, under grant 17716.1 PFES-ES.

Authors' addresses: M. Dolata, Department of Informatics, University of Zurich, Binzmuehlestrasse 14, 8050 Zurich, Switzerland, dolata@ifi.uzh.ch; D. Agotai, School of Engineering, University of Applied Sciences and Arts Northwestern Switzerland, Bahnhofstrasse 6, 5210 Windisch, Switzerland, doris.agotai@fhnw.ch; S. Schubiger, School of Engineering, University of Applied Sciences and Arts Northwestern Switzerland, Bahnhofstrasse 6, 5210 Windisch, Switzerland, simon.schubiger@fhnw.ch; G. Schwabe, Department of Informatics, University of Zurich, Binzmuehlestrasse 14, 8050 Zurich, Switzerland, schwabe@ifi.uzh.ch.

# 1 INTRODUCTION

When an advisor and an advisee engage in service interaction, they rely on a range of mutual expectations and obligations. Many of those expectations or obligations outgrow the institutional identities of the interactants [19], others result from the social character of the encounter [49,58]. In order to respond to the expectations and obligations, humans enact rituals. Some of the typical social rituals involve greetings, farewells, introductions [37]. Meanwhile, technology has become an integral part of many rituals. For instance, when agreeing for an appointment date, most humans pick up their mobile phone and it feels natural; many people even expect that others will use some tools when making an appointment. However, in face-to-face interaction, computers can also generate annoyance: consider the situation that occurs within service encounters, where the advisor is typing things on a keyboard and looks at the screen, while the client is sitting on the other side of the table and waiting; if this situation takes more time than the client expects, he[1] will likely get nervous or the advisor will engage in behaviors to bridge the waiting time [40]. In the best case, the situation will require a recovery behavior to put the interaction on the right track [13]. In the worst case, the high-touch, naturally unfolding encounter will turn into a technology-driven encounter, where the computer dictates behaviors and content [51,57]. In other words, the visible and explicit structure of information technology (IT) can dominate the structure of an interaction ritual. CSCW research provide evidence, that the interpersonal conversation and communication in professional encounters can become less effective and enjoyable due to IT [28,40–42,67]. The literature suggests a tension between using a computer during service encounter to save, process, or visualize the data (*high-tech*) and the ritualized, individual and human, character of the service encounter (*high-touch*) [2,74]. This paper questions this view and provides evidence that high-tech and high-touch in advisory services can be compatible. At the same time, it stresses the role of rituals, which should be considered a key source of inspiration and knowledge for the design of IT for institutional settings.

In financial advisory services, the high-touch rituals involve verbal as well as material and bodily interaction. Advisors employ pen and paper not because they lack alternatives but because pen and paper afford the high-touch interaction, which they use to impress the client [15]. Without those tools, some rituals like sorting and ordering paper on the table would be not possible, thus hindering the advisor at making the intended impression on the client. This manuscript describes LivePaper, a system that leverages existing pen-and-paper rituals and augments them with digital visualization and data processing. LivePaper goes even a step further: it enables new behaviors, which provide further ways to impress the client. The paper illustrates how the system gets integrated in the interactions between the advisors and their clients. The data collected through a post-treatment survey shows that the encounters supported with the system improve in terms of pragmatic quality and attractiveness compared to the traditional pen-and-paper services. We conclude that only by accommodating for interaction rituals, a system can combine the high-touch character of institutional encounters with the high-tech possibilities of modern IT.

This design study addresses the question on *how to enhance high-touch advisory services with technology.* It builds upon the claim that *combining high-touch and high-tech requires (1) support of physical rituals typical for advisory services and (2) implementation of further features that extend beyond those rituals and leverage the technical possibilities.* The results add to the previous body of knowledge on understanding and improving advisory services [15,16,23,28,32,42], as well as institutional interaction in general [1,22,51] being key fields of interest to many CSCW researchers. The results are relevant to designers and developers of collaborative applications for various application areas including business consulting, doctor-patient interaction, and service interaction. The insights have impact on the daily work of frontline employees, whose employers roll out IT support systems and applications. The ambition to streamline or improve the encounters often drives the organizations in a wrong direction – this manuscript shows how the goals of the organizations and the practical job understanding of their employees can be aligned in an effective design. The ritual-physical lens propagated in the current manuscript offers a theoretical fundament with a wide range of application areas. It is flexible enough to accommodate for the study of various co-located work environments, identification of requirements and potentials, as well as positioning of a system in relation to the physical configuration of routine work.

# 2 RELATED WORK

---

[1] To guarantee for an equal presence of both genders, while make the article accessible to the reader, we refer to the client as a male (he, his) and to the advisor as a female (she, her), thus implementing a suggestion from Pinker [60].

## 2.1 Advisory Services as Physical Rituals

To design effective support for advisory services, one needs to understand what an advisory service is. Advisory services have been approached by researchers from numerous areas. Management scientists study them as *dyadic decision situations* [36,55]. Marketing research sees advisory services as a customer relationship channel [46,61,65,73]. Literature concerned with supporting collaborative work frames advisory services as work [16,22,24]. Finally, sociologists and linguists see advisory services as a form of institutional talk and investigate how the sequential character of the services corresponds to the institutional identities of the participants [30,59,71]. The latter perspective includes recent works on the material aspects of the interaction between advisees and advisors [15,71]. Research offers multiple entry points to the study and framing of advisory services.

We conceptualize advisory services as work consisting of physical and ritual interaction between two individuals, an advisor and an advisee. This framing corresponds strongly to the institutional talk perspective [30,59,71] and to the socio-material lens [15,71]. In fact, the institutional talk discourse outgrows, among others, Goffman's research on rituals [19]. The understanding of rituals we leverage in here goes back to Collins [10,37], who adapts prior micro-sociological approaches, including the ones of Goffman and Durkheim. According to this concept, rituals require bodily co-presence, clear boundaries to the outer world, a mutual focus of attention, and shared mood as elements of a common event (including stereotyped formalities); through collective reinforcement, rituals produce symbols of social relationship, enhance emotions in individuals, provide standards of morality for a group, or strengthen the solidarity with other group members. Interaction rituals occur in mundane, daily interactions, they use and reinforce stereotypes of specific situations (which reside, e.g., in expectations on how an action should be conducted or responded to), and they address the ways in which members acknowledge each other as members of a group in their specific roles [10,37].

There are two central implications of using rituals as a lens: First, interaction rituals emphasize *situation* over the individual who take part in the situation ("Not then, men and their moments. Rather, moment and their men." [25]) – this contrasts the ritual perspective to the practice lens, which emphasizes structures outside the particular situation, such as the individual experience or the organizational context of the situation [56,64]. Of course, rituals could not emerge without a common mood or understanding of stereotypes, but the ritual lens directs the focus towards the situation which drives the behaviors of the participants and adds to their experience rather than the other way around [10]. Second, the rationale behind acting the one or the other way does not emerge through balancing out external structures, but it depends strongly on how a participant wants to present herself during the interaction and what face she needs to maintain during the interaction. Therefore, the reasons for specific actions of an individual reside within the situation – in previous actions – and not outside. Although, it is clear that the context might limit the choice of what role each participant might take. So far, the ritual perspective as proposed in the current paper has been only rarely considered in the CSCW and then, mostly, in theoretical considerations [63], however, in related disciplines it has been recently employed, e.g., in the context of game design [5,21].

Through explicitly referring to rituals, we want to stress that the performances during advisory services (1) have a routine character following specific stereotypes, (2) are meaningful only in the particular situation and emerge as consequence of previous performances in this situation, and (3) celebrate a specific order, in which the participants act to establish or protect their reputation [10,25,37]. Previous research explicates that the advisor during an advisory encounter establishes and protects her identity as a domain expert, a representative of an institution, and a customer-centric frontline employee [16] and that seemingly independent actions form a whole in this effort. By emphasizing the physical character, we point out, that the space, human bodies, furniture, and materials all belong to the ritual [18]. In *physical rituals,* humans act according to obligations to others and expectations one has of others while employing symbolic acts, which involve material and space [31]. For instance, advisors and advisees distribute documents in space to illustrate complex matters and to exercise control over interaction space, use expressive gestures to introduce oneself, employ posture to signalize attention or disinterest, and take care of the surrounding during an advisory service [15]. In mundane actions the physical and ritual nature of advisory services comes to the surface. Designing a system for advisory services requires careful handling of the ritualized physicality.

The holistic lens applied in the current study acknowledges the fact that material, the area accessible to the interactants, and the physical environment surrounding them, all form the interaction: on the one hand, they afford specific performances while limiting others; on the other hand, they have been designed with specific routines and rituals in mind. This inherent relation between advisory services and the characteristics of the physical circumstance are coupled so tightly that the rituals' occurring depends on the place (rooms, locations, buildings), conditions (light, furniture, sitting positions) and specific elements (documents, sheets of papers, pen) [15,18,71]. Figure 1 shows the physical configuration by illustrating the relationships between the physical areas. Note, that the particular areas correspond to Kendon's o-space (interaction area), p-space (access area), and r-space (environment) [38,39]. However, to differentiate from Kendon, who studied how the spaces and new formations between individuals emerge, we prefer to stick to the more general term *area*, which also better expresses our interest into what happens within the area [8]. In the current study we treat the physical configuration of advisory services as a way to frame and organize the requirements for a system to be developed.

Apart from spatial and physical configuration of advisory services, existing practices can provide knowledge about the advisory service ritual and generate requirements for a system too. In particular, when attending to an advisory service, e.g., at a bank, advisees expect not only a clean and tidy room, and an advisor dressed in a specific manner, but they expect that the advisor will take notes, use tools to make calculations, and refer to leaflets and official documents [15]. The expectations from the one side match the performances of the other side. Many bank advisors rely on an advisory paradigm called *pencil selling*. It involves use of empty sheets of paper and a pen to illustrate what is being said with ad-hoc drawn pictures, calculations or text [20,69,70]. The pencil-selling practices all involve writing and drawing on paper, moving and freely positioning paper within the interaction area, as well as removing it from the interaction area and piling it on the side [15]. Overall, the *pencil selling* method – even though being often popularized as a specific marketing concept – describes the advisory ritual in terms of physical conduct [15].

Pencil selling dominates financial advisory services, but things are changing. First, after the crisis of 2008, banks are increasingly responsible for educating their clients and documenting this education effort – drawings or calculations may be insufficient [55]. Second, financial products are increasingly complex, such that the drawings and explanations based on brochures do not hold on to this complexity [54,55]. Third, some advisors who feel uneasy at drawing turn pencil selling into calculation exercise: the advisor calculates values in front of a waiting client – be it on computer or on a calculator [15]. This makes the face-to-face services unattractive to many clients who may then switch to robo-advisors [3]. We argue, that the face-to-face services can benefit from a smart combination of human advisor and IT, such that the IT complements the human by contributing skills in which computers are strong (e.g., visualization, calculation) and vice versa [7]. The human advisors are strong in communicating and explaining complex products by using simple and individualized information and by presenting it in a way that fits the particular client [15]. They are also experts in rapport building [67] and impression management [15]. No computer can take over those tasks easily. However, previous research on IT for advisory services often disregarded the advisors' soft skills: IT imposed a specific order of actions [53] or forced the advisor to explicate information about the advisee to enable for further processing [41,42]. And, what is more relevant, the prior solutions took the basic tools, pen and paper, away from the advisor. The physical rituals were limited, such that advisory services did not unfold in line with the social expectations and obligations. This study aims at complementing advisors' material, ritual practices with IT and affording new compatible performances. Accordingly, it relies on requirements we derive from the current physical configuration and practice.

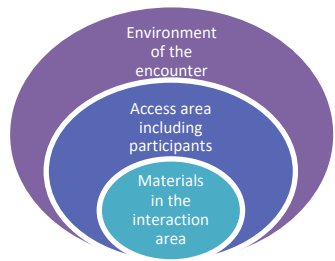

**Fig. 1. Physical configuration of advisory services**

## 2.2 Challenges of a High-Touch Service

A collaborative application needs to respect several requirements to enable for a high-touch service. Some requirements follow from the routine behaviors in which advisors engage during the service. Some are a consequence of the physical characteristics of the environment or embodied interaction expected by the clients. In particular, to engage in pencil-selling practices, the advisor should be able to handle paper the same way, she does it during conventional services. The advisors use paper to establish common focus (by positioning a single sheet of paper in the interaction space), to compare options (by positioning several sheets next to each other) or to introduce new information (by putting a sheet of paper into the interaction space). Dolata and Schwabe [15] provide a list of typical paper performances that support the advisory service as a ritual. Table 1, rows 1-3, summarize the requirements that follow from pencil selling as the main interaction paradigm. Pencil selling requires specific environment and circumstances to emerge: for instance, it works much better on a table (which establishes a natural interaction space), in a bright room (such that participants see what's written on paper) and a tidy environment (which reduces the distractions). This may seem obvious at first sight, but some modern technologies using projection or built-in screens fail to function properly in bright rooms or may be bulky, difficult to hide and clutter up the interaction space. Therefore, we capture the requirements which assure the right space and environment for the high-touch service to emerge in Table 1, rows 4-6. The requirements rely on the assumption that a high-touch service emerges when expectations and obligations concerning the ritual conduct and the physical space are fulfilled. However, where is the point of building a system that solely digitalizes what is already there, in place? We argue that modern collaborative technologies can generate value for advisory services which cannot be achieved through traditional pencil-selling approach.

**Table 1. Requirements to support high-touch services according to the current practice and setting**

| Design Requirement | Physical scope |
|---|---|
| 1. Support physical handling of paper such as free, dynamic positioning | Interaction area |
| 2. Support creation of individualized content through handwriting | Interaction area |
| 3. Support introduction of new material into the interaction area | Interaction area |
| 4. Enable advisor and advisee to access to the shared interaction area | Access area |
| 5. Enable storing of unused material on the table, outside the interaction area | Access area |
| 6. Enable interaction in a bright room furnished in a standard way | Environment |

## 2.3 Potentials of a High-Tech Solution

A dedicated collaborative application can enhance the quality of advisory services in many aspects. Previous research documented improvements in terms of transparency [11,53], in-situ customer education [29], documentation [23], access to data [22] or persuasion [12,16]. Whereas the studies offer specific design principles that fit their individual goal, they share central concepts. The transfer of messages from the advisor to the advisee enhances when the advisor can use multimedia and multimodal representations during the service [22,23]. Complex relationships between multiple relevant factors can be better understood by clients if they can observe (and thus experience) how things depend on each other when changes occur [22,29]. And reusing the same, recognizable content (e.g., the same visualizations) accentuates the central message and supports remembering it [12,16], as well as makes the service overall easier to follow [11,53]. Overall, the studies advocate (1) a multimodal approach to engage multiple senses during an advisory service [4,45], (2) experiential learning to make complex relationships cognitively accessible [43,44], (3) and the object constancy to make the advisory service traceable [47,62,72].

The multimodal approach towards communication claims that providing various sensory information during communication provides a better access to its content and reduces cognitive load related to understanding the message [4,45] (unless the information provided through sensory channels is inconsistent). Pencil-selling uses verbal and two-dimensional visual cues to transfer the message. Mixed reality approaches, like spatial augmented reality [6] or tangible interfaces [33,34] incorporate physical objects and surfaces into digital interaction to engage spatial perception and allow for touch as a sensory cue. However, the solutions for advisory services were limited so far to simple graphics, diagrams [22], and video or audio [12]. It remains open how tangible elements can co-exist with pencil-selling practices.

The experiential learning paradigm asserts that humans learn easier through experiencing how things depend on each other rather than through being confronted with theoretical accounts describing the relationships [44]. Whereas humans can easily imagine linear relationship based on a theoretical explanation, irregular or exponential relationships demand much cognitive effort [7]. Therefore, it is increasingly popular to make users (pupils, museum visitors) engage with physics laws, chemical rules or statistics by making them experiment and change interdependent values [43,44]. Pencil selling is limited in terms of experiential learning: advisors draw how financial factors (such as the price of a property and the height of a mortgage) are related to each other but lack tools to represent this relationship in a dynamic way. Previous research in this context shows that dynamic experimentation with numbers can support a layperson at making more informed decision but those system limit the experiential learning to dedicated episodes in advisory services, the so called microworlds [29], thus decoupling the experiential learning from the actual, individual advisee's case. It remains open, how experiential learning can be integrated in a financial advisory service.

Finally, the object constancy (also called object permanence) is a concept from development psychology that describes how humans learn to recognize an object occurring at two different time points during an interaction to be a single entity [47,62,72]. It is an important step in children's development to learn that a ball, which the mother is hiding behind her back, does not disappear for the moment and re-appears again later on, but continues to exist. Having learned this principle for physical objects, we can rely on it in many circumstances. The same holds for pencil-selling encounters: when an advisor writes something on paper, then puts it away for a certain time, and then puts it back again in the middle of the table, the client will recognize the sheet of paper. Even though he might not have tracked the paper all the time, he will be able to comprehend the information and put it in context. For IT-supported advisory service, it was shown that taking pictures of specific artefacts makes it significantly easier for the advisees to recall abstract information concerning this artefact or related issues than an exact description of the artefact [23]. Also, visualizing focus changes in a software for financial advisory services helped the advisees keep track of the advisory process and outperformed an agenda list [53]. However, object constancy contradicts the experiential learning and multimodal approach paradigms: how to make things constant while allowing for manipulation of values? And how to integrate it into a system that allows for free movement and handling of paper?

The current study envisions a system that combines multimodal content, experiential learning and object constancy with the traditional pencil-selling approach. The envisioned system shall contribute to the encounter's quality and, at least, preserve its high-touch components. Table 2 lists the potentials for improving the financial advisory services while formulating them as requirements to the envisioned system. Whereas the potentials have been identified in earlier research, the systems implemented them often in limited manners (e.g., in specific episodes, without incorporating spatial or motoric cues). Additionally, those studies tried to replace advisors' practices rather than preserving the rituals, thus risking disturbance to the high-touch character [28,29]. Overall, even though previous research points to specific potentials of collaborative technologies for advisory services in banks, police, healthcare or energy saving, the community still lacks guidance on how to make the interaction high-touch and high-tech at once. As a consequence, the design efforts of practitioners lead to ineffective and unpopular solutions. Deployed and put into daily business, the systems then disturb the ritualized conduct between the interactants and thus generate more harm than good, as illustrated by numerous field reports [16,40,51,57].

**Table 2. Requirements to improve advisory services through high-tech interventions**

| Design Requirement | Concept |
|---|---|
| 7. Introduce physical objects to engage spatial and sensory perceptions | Multimodality |
| 8. Introduce dynamic graphics to make abstract relations comprehensible | Experiential Learning |
| 9. Keep objects and their content stable and recognizable, while allowing for manipulation and movement to support object constancy | Object Constancy |

## 3  Design and Evaluation

### 3.1  LivePaper = High-Touch + High-Tech

Pencil selling involves extensive and diverse use of paper and printed materials between the advisor and the advisee [15]. The LivePaper system does not intend to change the basic interaction principles, instead it enlivens the paper: the medium which has been used for interaction between the advisor and the advisee gets augmented in a way that complements existing tools with computer-generated visualizations. This allows for new interaction mechanisms to support multimodality, dynamic graphics to support experiential learning, and establishes a constant link between a sheet of paper and its content, such that object constancy can be guaranteed. The environment is a normal advisory service room – in the room's ceiling we integrated a projector (Optoma UHD60 4K) and sensors (Microsoft Kinect) necessary to control the interaction in the interaction area. To accommodate for natural light conditions, we chose an ultra-bright overhead projector and used high-contrast reflective markers on the tokens and on the sheets of papers. The access area consists of a table, two chairs at opposite sides of the table, and the bodies of the participants – to ensure good projection quality we used a standard white table; otherwise no adaptations were necessary. The interaction area emerges on the table, between the two participants sitting in their chairs. The area is approx. 1.5 m², which is around ½ of the table. Within this area, the user can employ the developed interaction elements which respond to the identified requirements (see Fig. 2).

The projector and sensors are calibrated to illuminate and observe this area only. As presented in Fig. 3, the interaction space includes materials used by the advisor or the advisee: they may include plain paper, plain paper with dedicate markers (recognized by the system), a notepad and a pen, brochures, and other documents. Additionally, the system provides tokens recognized by the systems, which activate specific functions when placed in the interaction area. The system projects content only when a token or a marked paper is within the interaction space and the illuminated area is limited: the graphics directly surrounds the token or is limited by the size of the paper. Consequently, the system does not make essential changes to the physical configuration. If the participants do not introduce the tokens or marked paper into the interaction area, the system remains passive such that no change can be noticed.

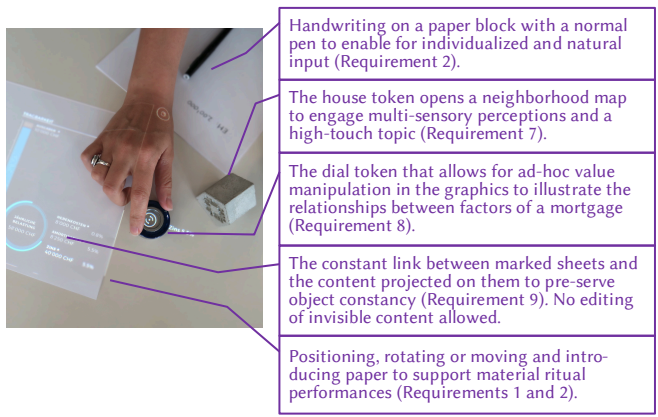

Handwriting on a paper block with a normal pen to enable for individualized and natural input (Requirement 2).

The house token opens a neighborhood map to engage multi-sensory perceptions and a high-touch topic (Requirement 7).

The dial token that allows for ad-hoc value manipulation in the graphics to illustrate the relationships between factors of a mortgage (Requirement 8).

The constant link between marked sheets and the content projected on them to pre-serve object constancy (Requirement 9). No editing of invisible content allowed.

Positioning, rotating or moving and introducing paper to support material ritual performances (Requirements 1 and 2).

**Fig. 2. Design intentions affecting the interaction area and their relation to the design requirements**

The interaction with the system is possible through a range of channels. First, the system recognizes handwriting and drawings on the notepad. It uses the Wacom Bamboo Slate to capture strokes on paper, which are then processed by the Microsoft Windows' built-in handwriting recognition module. The system monitors for specific code words to identify relevant values – the current list of code words mirrors the advisors' shortcuts used during conventional advisory services (e.g., "*EM*" for *Eigenmittel*, German for *net assets*). Second, the Microsoft Kinect sensor provides information on user actions applied to the content projected on the table (e.g., clicking or marking an area), as well as actions applied to the tokens or paper (e.g., position change, removing from interaction area). Recognizing the position and movement of marked paper or tokens is essential, such that the projection can follow the paper or token and thus establish the illusion of object constancy. The individual marking allows for constant link between specific content and a marked sheet of paper or token. Third, tokens allow for specific inputs as well. In particular, one token which works like a volume dial and allows for continuous change in values projected on paper Fig. 4 – left). When values are adjusted this way, the projection adapts dynamically such that effects of changes are immediately visible. This supports experiential learning: applying changes to one variable (e.g., monthly interest rate) causes visualized adaptation of other variables as well (e.g., overall price of the mortgage, monthly budget, or duration till full amortization). Other tokens (Fig. 4 – right), which resemble square gambling chips, stand for tranches (portions) of a mortgage. By adding them to a loan, the client may change the mixture of risks (flexible and stable interest rates) – they support spatial perception and provide a direct, multimodal impression of how volatile or fixed the rates are. The third token, a house, opens a map of the neighborhood in which the desired property is located. The three-dimensional house token in the projected landscape activates the multi-sensory impression and, simultaneously, introduces an emotional, high-touch topic into the conversation. Overall, the system provides means to involve multimodal communication and experiential learning, while leveraging the constant binding between physical objects (tokens and paper) and content, thus making way for paper practices (e.g., moving, positioning, or piling) known from pencil-selling. We encourage the reader to view a video of the system[2].

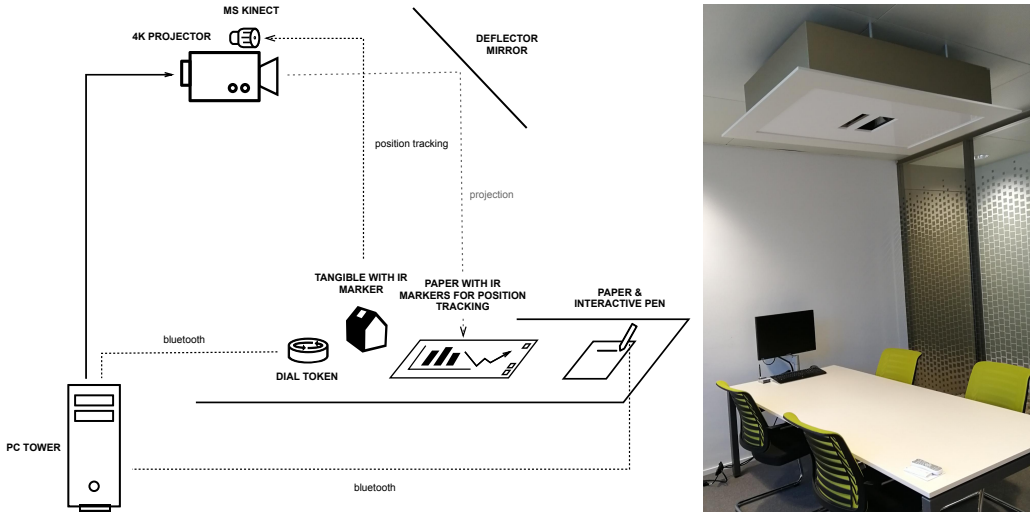

Fig. 3. Left: Technical setup of the LivePaper. Right: The system hidden in a box at the ceiling.

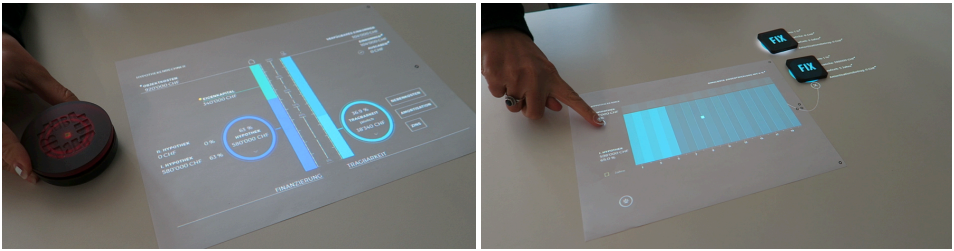

Fig. 4. Left: dial token in use to adjust the income.
Right: a second fixed tranche being added to a loan.

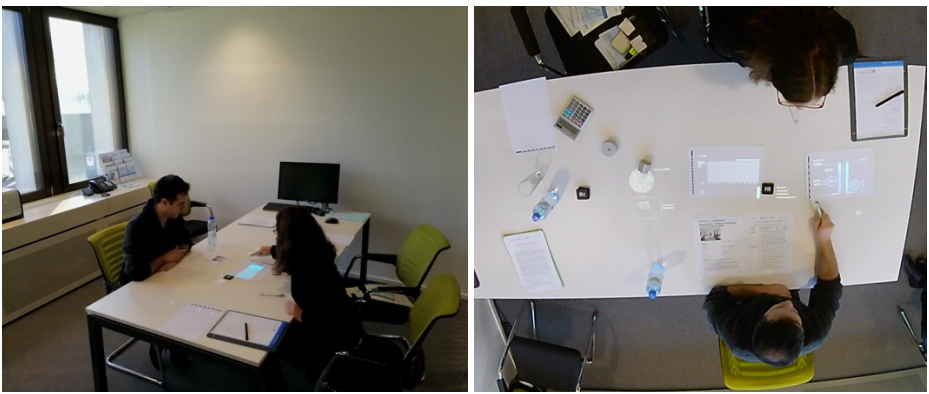

Fig. 5. The system in use during an advisory service: Left: side view. Right: top view.

---

[2] https://www.ifi.uzh.ch/en/imrg/research/advisory-service-support/live-paper.html

Accepted for publication in PACMHCI

The system provides a range of features to support an advisory service. They were intended to be used as follows. An advisor, let us call her Ann, has an appointment with a new client, Clark. Upon initial small talk, Clark makes explicit, that he wishes to buy a new house for his family. He has brought some basic data about a property he has visited shortly with a real estate agent. Ann uses the interactive pen and paper to note down some data like the address and the price. While Clark is explaining his plans concerning the house, Ann puts the house token in the middle of the table: a map appears showing the location of the property. Ann and Clark look at the map and Ann recognizes that she knows the neighborhood and reminds herself that there is a family restaurant near there. She uses the dial token to zoom in and by clicking on an icon left to the map, she switches to the satellite view. She now can identify the restaurant based on the sunshades outside of a building. It is only 200 meters away from Clarks future property. They continue talking about the neighborhood in general and about the size of the parcel size that surrounds the house Clarks wishes to buy. By "clicking" another icon, Ann minimizes the map and moves the house token away. Clark and Ann need to consider the finances. Ann puts a marked sheet of paper in the middle, she clicks on a label saying "Affordability", and a graphics appears on the paper. At first, an area labelled "income" shows red: Clark alone would not be able to fulfill formal requirements, but Ann and Clark quickly realize that they have not yet considered the income of Clark's wife – Ann clicks on the label and uses the dial token to adjust the values accordingly (cf. Fig. 4 – left). Now the graphic is all blue – a good sign showing that the loan and the house are affordable for Clark's family. Clark is happy. Ann offers Clark to look at how much an actual loan would cost and puts yet another paper in the middle of the table. She clicks the label "Motgage Mix" and positions a chip "FIX" next to it explaining what a fixed mortgage means. She adds the first fixed tranche to the mix by clicking a "+" icon (cf. Fig. 5 – left) and then a second one (cf. Fig. 4 – right). By using the dial token, Ann adjusts the duration and the size of the tranches while discussing the various options, their advantages, disadvantages and their impact on the monthly rate. From time to time, Clark asks questions while referring to the graphics on the table (cf. Fig. 5 – right). Overall, at the end of the encounter, Clarks knows that he can afford a property and has understanding of his monthly expenses related to it.

## 3.2 Evaluation

The evaluation was conducted to understand whether and how LivePaper can combine and support the high-quality and high-touch dimensions. Consequently, it was of great importance to (a) create situations where LivePaper can be used and compared with the traditional service, (b) measure the differences concerning the overall quality and high-touch features between the two service types, and (c) observe what behaviors changed due to LivePaper to better understand why the effects occur. To balance out the trade-off between the reality of the situation and the possibilities to make extended observations and collect opinions concerning the quality measures, we decided to employ an instance of LivePaper in a local bank and run an experimental series of realistic advisory services with real advisors and test advisees.

The LivePaper was instantiated with all features presented above to support mortgage advice services at a regional, Swiss bank, MoBa, with 13 branches and approx. 400 employees, who serve an area inhabited by approx. 600,000 people. MoBa's advisors provided regular feedback during workshops and formative tests. The evaluation in the current study tested the first functional version of LivePaper in a MoBa branch in January and February 2017 and was arranged as a within-subject design experiment [48,75]. Six selected advisors, others than those who participated in the design process, were chosen for this evaluation. They received half-day training on the usage of LivePaper – they learned about the system functionalities and simulated an advisory service with a colleague. During the evaluation, each advisor advised three different test advisees while providing one conventional and one LivePaper advisory service to each of them. To balance out the order effects, we varied the treatments order. The 18 test persons acting as advisees were acquired through official advertisement website of the University, available to the broader public and linked with social media. They were offered a compensation of 60 CHF (approx. 60 USD) for 2.5-hours experiment. On average, the test advisees were 27.5 years old, with the youngest participant aged 20 and the oldest one - 49. Their professions included shop assistant, nursery teacher, designer, or veterinarian assistant. Only three test advisees declared that they attended to a face-to-face financial advisory service before, but 10 of them at least considered going to a real advisory service in the near future. Accordingly, the chosen group reflected the standard knowledge and experience level of clients attending to a mortgage advisory service when buying a new property – normally only few would have previous experience. The test clients were provided a rough handout including information they could use as inspiration when acting as an advisee: a hypothetical financial situation, the property they pretend to buy, and a few questions they could ask. However, the advisees and the advisors were free during the encounter – some conversations deviated from the handout depending on the interest. After the experiment, the advisors and the advisees assessed the overall atmosphere to resemble real advisory services. By involving real advisors and real environment in a bank branch, and by designing the whole experience to be as realistic as possible (e.g., with only very rough scenarios such that test clients can possibly act in their own name) we intended to take care of the external validity of the current study. This was of high importance, such that the advisors can assess the applicability of the system in their daily work and the MoBa could make an informed decision on the further course.

The experiment enabled for the collection of a range of data. First of all, all advisory services were video-recorded. In the months after the experiment, they were transcribed according to the Jeffersonian notation [35] with extended comments concerning material and bodily behaviors. All transcriptions were carefully reviewed by two researchers to identify patterns of behavior occurring across the whole data set. The identified patterns were discussed in workshops with advisory service experts from the bank to establish a consistent and most adequate understanding across the research team. In particular, we identified (1) performances that resemble pencil-selling practices, (2) performances that originate at pencil-selling but incorporate new digital elements or are adapted to them, and (3) performances induced by the LivePaper. We focused on whether new or strongly altered performances generate ineffective communication patterns (silence, breaks, unfilled pauses, missing eye contact), which signalize that some expectations or obligations typical for the ritual are not fulfilled. Overall, video analysis identified design elements with negative or positive impact on the high-touch interaction. The authors identified representative excerpts for each of the identified patterns involving the use of design elements to be included in the current manuscript, such that an overarching picture of a LivePaper-supported advisory service emerges from the provided data.

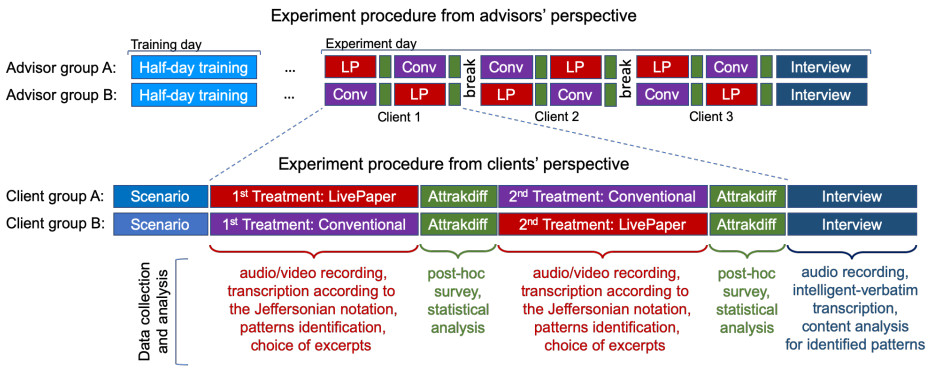

Fig. 6. Experiment procedures from advisors' and clients' perspectives with data collection and analysis activities during and after the experiment.

Second, the experiment yielded post-hoc data concerning the advisory service quality. Each advisee and each advisor participated in a semi-structured interview of approx. 30 (advisees) to 60 (advisors) minutes after all their treatments. The interviews were structured around the experiences of the participants (most enjoyable, least enjoyable episode from the advisory services), as well as the overall impression of the advisory service and its particular phases. In 2018, after the transcription and analysis of videos, a single researcher analyzed the interviews – he identified passages from the interviews that referred to the patterns and performances identified in the videos and those providing an overall assessment of the experience. This allows for the appraisal of the emerging rituals from the perspective of the participants. Apart from the interviews, directly after each treatment, we administered a short questionnaire to each participant. To obtain a holistic comparison of different aspects of the services, we employed the Attrakdiff [26,27] instrument. It captures the pragmatic quality (related to qualities like practical, structured, simple) as well as the hedonic qualities (identity and stimulation; related to qualities like exciting, inclusive, classy), and finally the general attractiveness (e.g., inviting, appealing, pleasant) [27]. The instrument relies on differentials and asks the respondents to choose between opposite adjectives (e.g., confusing and clear) on a 5-point Likert scale and has been previously applied in research on advisory services [52,66,68]. We see the pragmatic quality as a dimension related more to the high-tech aspects – it results from the user assessing such aspects as *simple vs. complicated* or *practical vs. impractical*, while hedonic qualities relate more to the high-touch aspects of advisory services – they relate to aspects like *separates me vs. brings me closer* or *dull* *vs. captivating*. Attrakdiff sees attractiveness as dependent on the pragmatic and hedonic qualities [26,27]. The analysis of multiple data sources indicates (a) whether the design objectives of making the encounters high-quality and high-touch were reached and (b) how they were reached by showing what performances emerged at the intersection of the new technology and the ritualized performance.

# 4 RESULTS

## 4.1 Overall Assessment

Attrakdiff results make clear that LivePaper advisory services outperform conventional ones in terms of high-touch and high-quality aspects of an advisory encounter. Further studies suggest that the LivePaper has additional advantages concerning marketing and information provision aspects but referring to them in this manuscript would move its scope far beyond the high-touch and high-tech issues. See [Anonymous – under revision in a marketing outlet] for further results. Concerning the high-touch and high-tech aspects, advisees experience enhancement concerning the pragmatic quality (t(17)=3.69, p<0.005), as well as the emotionally loaded qualities: identity (hedonic quality; t(17)=2.40, p<0.05)) and stimulation (hedonic quality; t(17)=5.97, p<0.001). The general attractiveness is significantly higher too (t(17)=4.07; p<0.005). Advisors experience improvement on the high-touch qualities, identity (t(17)=4.05, p<0.001) and stimulation (t(17)=3.41, p<0.005), and attractiveness (t(17)=2.74, p<0.01). No noteworthy difference could be observed between the first, the second, and the third LivePaper treatment. Overall, clients notice improvement in all dimensions, while the advisors experience improvement in the emotional attributes.

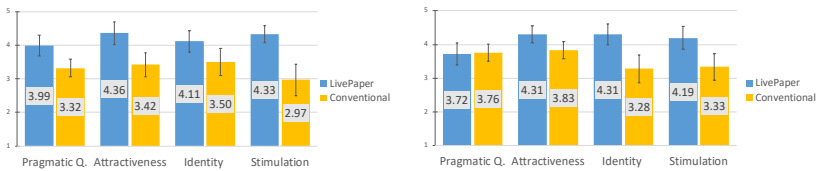

**Fig. 7. Attrakdiff [27] assessments: Left: advisees; Right: advisors; (min. 1 and max. 5; error bars: 95% CI).**

## 4.2 Performances and Opinions

The current section provides a chronological blow-by-blow analysis of a single experimental advisory service. The section refers to features implemented in LivePaper (cf. Figure 2). We select a number of excerpts that illustrate typical, repeating patterns of how features of LivePaper were embedded in the encounters by the participants. The numbers in brackets (1, 2, etc.) refer to the line numbers in the transcription. The original Swiss German transcription was translated into English for this article. To make the transcription both rich and accessible to an international audience, we use normalized English (e.g., `you` instead of `u:`) but take care for maintaining general prosodic signals: `[` stands for overlapping speech, `:` signalizes a long vowel, `[.]` and `[X.Y]` stand for silent breaks and pauses, `CAPITALS` indicate louder syllables or words, `°` indicates quiet or softly-spoken words, `/` and `\` indicate rising and falling pitch. `A` refers to the advisor and `C` to the client. The non-verbal behaviors are in **bold**. For the interpretation of the observations, we refer to the interviews conducted with the advisees and advisors.

*4.2.1 Pen input.* We join the advisor (A) and the client (C) right at the beginning of a service. They are seating at a table, opposite to each other. The atmosphere is relaxed, oriented towards the service. From the very beginning, A signalizes the intention to come to the merits – verbally (line 1) and non-verbally (2, 5).

```
1   A: so: best welcome [.] nice you have found the time today [.] a:nd lets directly begin [.]
2      A picks up her pen and the notepad, C looks at A
3   C: °yeah°
4   A: so: tell me a bit what brings you he:re/
5      A puts the calculator away, places the notepad in front of her, holds up the pen and opens it
6   C: so i am looking for [.] I me:an ive got an object [.] I could [.] could simply show [.]
7      C picks up his folder and looks through documents inside
8   A: yes su:re [0.3]
9   C: A::nd eghm [1.3] yes its a: [0.6] a flat here in lenzburg and just the question of financing it
```

```
10       A looks at her notepad, C takes out two documents out of his folder and places one of them, a flat
         data sheet in front of the A, while puts his list of notes to his left
11  A:   mhm
12       A looks at the data sheet and takes some notes
13  C:   and eghm [.] yeah [0.5] and i am employ [.] eghm [.] i am currently EMPLOyed
14       C moves his notes away, A continues taking notes
15  A:   YEAH
16  C:   and I could get inheritance advance from my parents
17       C looks up to the A, A holds the notepad and nods
18  C:   quite high [.] I mean [.] maximal height of two hundred two thou:sand francs
19       A continues taking notes on the notepad, C looks first at the notepad than at the A, A gazes at C
20  A:   okay: [0.3] so: you mean its lenzburg huh
21       A looks at the notepad again, C looks at A
22  C:   °ya° [°exact° (0.3)
23  A:           [°now is° lets try (0.5) FLO:RA:street si:x [.] huh/
25       A writes "STR = " and then points with her finger to the data sheet and looks up to the C, C gazes
         on the notepad
25  C:   [exactly]
26  A:   [THIS is] the PROPerty/ [2.2)
27  A:   °okay°
28       A moves the notepad slightly to the centre of the table, A notes down "STR = Florastreet 6"
29  A:   °so I will note this down right° (5.5) so:: [2.5)
```

We observe the beginning of the encounter – shortly after the greeting, the advisor starts taking notes (12). When noting down the street name, she says it out loudly (23) and later explicates that she makes the note of that, while taking her time to finish writing down (29). Through this behavior, A puts strong focus on the activity of note taking – stronger than observed in accounts of conventional note taking [50]. Additionally, long unfilled pauses (29) as well as signals of uncertainty ("let's check", 23) show that the note taking is special. Based on this action, the client could have identified that the advisor produces notes in an outstandingly careful way, such that the system can recognize them. Humans use note taking not only for their own use – it is a typical sign of careful attention paid to the opposite party, such that the action can be seen in terms of rituals. Whenever one provides much relevant information in a condensed manner, note taking as a situation seems to be natural. However, the situation we observe above is an exaggeration. Because LivePaper uses notes taken by the advisor for further processing, the advisor makes additional effort to produce exact notes in a computer-readable manner. Even though the designers hoped for the interaction with the interactive pen and paper to be more seamless, the episode shows clearly what challenges occur when high-touch meets high-tech – in few cases the high-touch aspects might get dominated by high-tech.

Opinions from the clients and the advisors provide a range of interpretations. Many clients did not notice any difference. *"It was natural. The pen looked quite normal. The paper was simply a standard paper. Nothing special, I could notice."* Some clients referred to the amount of information collected by the advisor as being lower in LivePaper: *"I found it okay how she was making notes. In the traditional setting, she wrote down quite a bit of information, while with LivePaper she just made the most essential notes but left out some other elements (…) [Writing down much information] was a signal to me, that she wants to collect lots of data, to give me a sound advice. This goes down a bit with LivePaper"*. Another topic addressed in this context was the readability of notes: *"In the conventional setting, I could barely read and control what she wrote down – too small, too far away. In the LivePaper, her handwriting was clear and tidy, such that I was sure she would have the right data about me"*. Overall, the clients focus on the output rather than on the process of taking notes and accept such behaviors as spelling the syllables as casual. Advisors take more direct reference to the interaction during note taking. Especially those who experienced problems, because the system did not recognize the input, reported: *"Well, it's good and helpful, but if you need to write down the number three times and it doesn't show it recognizes the input, then you might be a bit distracted. But I wasn't distracted. (…) But the client sees that you write and it doesn't work, so he would wait and not talk. So even if you are receptive to his talk, it's his assumption, you're not. But this could also happen when you're doing calculations on paper"*. Others see a great advantage concerning the efficiency: *"I like the whole. So to say, I write down the equity, own funds and so on on the block, and it takes the entire data into the system. That is very nice and takes less time. In other words, if you do not have to use the keyboard type in the data, you can really and directly work with the visualization and so on."* Overall, the pen-based input in LivePaper may generate phases of unnatural interaction, e.g., when advisor repeats or spells out words, but this does not cause critical communication breakdowns – neither advisors nor client find this particularly distracting.

*4.2.2 House token.* Having collected the address and key data, A and C continue with the most emotional topic in mortgage advisory services – the house. Conventionally, the advisor would shortly discuss or note down the basic data of the flat, such as the number of rooms, size and additional features. This information has pragmatic relevance, which the bank needs to decide about the maximal sum of the mortgage. LivePaper provides the house token and a map to extend the range of modes to introduce and address the property – this excerpt shows this interaction.

```
66  A: first [.] id say [.] we look at the flat
67  C: ˚mhm˚
68  A: [lets say weve got a ma:p [5.0]
69     A picks up the house token from her left and places it in the middle of the table, between her and
       the C.
70  A: ˚sure˚ [.] we can quickly ahm to the side [.] have a loo:k together .h
71     A picks up the pen and points with it to the numbers positioned to the map's right side, C follows
       with gaze
72  A: here weve got everything [.] so: for you so that we: ca:n check the correctness [.] we have
       PURCHASE PRI:CE seven hundred seventy thou:sand
73     A points with the pen to the numbers, C bends over slightly and looks at this position
74  A: .h eghm: florastreet six in lenzburg
75  C: ˚mhm˚
76  A: it looks okay to me like tha:t [.]
```

The introduction of the property is made verbally explicit (66) and occurs simultaneously to the introduction of the house token to the interaction area (69). Thereafter, A uses a summary screen positioned aside of the projected map to review the collected data (72, 74, 80). A clearly dominates the stage in this interaction – C's contributions are limited to "mhm" (67, 75). Thereafter, A and C discuss C's current living situation without using the map, the house or the dial token, but those three elements remain in the interaction area. Having finished the exchange, A and C move their attention (and gaze) to the projection and tokens on the table.

```
92  A: ˚then lets have a look together˚ [0.9] ˚so we:˚ [.] so that YOU can MAKE YOURself a picture [1.8]
93     A dials on the dial token, zooms the map in, C looks at A and her dialing the token, then A and C
       gaze on the map
94  A: maybe a few CORNER points you sure know [0.4] ZURich
95     A points with his pen to the respective area of the map, C looks at the map and follows the pen
       with his gaze
96  C: ˚mhm˚
97  A: we have here ZURich [.] we have down here have already lucerne lake lucerne [0.8] ah: BAsel/ [.]
       [A's long turn omitted for space reasons] do you have a car/
98     A talks slowly and points to the respective areas of the map with her pen, looks up to C
       sporadically, towards the end she gazes at C, C bends down towards map and gazes the whole time at
       the A's pen, then looks up to A
99  C: i have currently no car because i want to tra:vel with public transportation [0.5] .h
100    C and A look at each other
101 A: [mhm
102 C: [and a:h the connection [0.2] i would not know how it is ˚honestly˚
103 A: so: i show this to you now [0.2] the bus stop is [1.2] is not far away actually [1.6] ˚i will go a
       bit narrower˚ [1.3] i will zoom in [2.4] and click the map [4.5] so: here is the house and here is
       the bus stop
104    A rotates the dial token several times to get down to the proper zoom level and clicks to change
       the mode to satellite, then A points with her pen to the map with her pen and then to a marking on the
       map, C bends over to the map and looks at it carefully, then C points to an area
105 C: COOL [0.9] and theres a ˚green path˚ there:/
```

The discussion about the location has two parts: one concerned with the location in relation to other cities (97) and one concerned with the house in relation to a point of interest (103). Whereas the long A's monologue on the general town location does not generate active reaction from C, except from passive observation (98), the zoom-in generates a positive, emotional reaction (105). The advisor stresses the fact that the activity they both engage in is oriented towards the advisee (92) – since the sentence is a self-correction (we vs. you), this effect gets even further amplified. Making the beneficiary of the ongoing interaction explicit helps the advisor showing her dedication for the client and, thus, responding to the potential expectation of a client-centered advisory service. Interestingly, what happens in the interaction area (dialing the token, changes in the visualization because of the zooming in) can even further amplify the impression on the advisee, because he clearly sees that there is "work" going on. He, finally, explicates what impression the visualization makes on him with an emotional statement (additional to several "mhm").

The advisees agree, that visualizing the map and discussing the house location does not form a central element of an advisory service and consider this a gimmick. However, they differ to how they assess this gimmick – as an enrichment or as an unnecessary bauble. Discussing the property and its features seems interesting and relevant: *"So first of all the beginning with the small house. This addresses more the emotional level. Because you want to buy an object, and yes, then you can see it again on the map. That's more like a gimmick, but I think that's okay, because it makes you emotionally involved."* Some clients even expect a re-assurance from the advisor despite her true opinion: *"I think that's a nice start, that's the icebreaker too. Maybe then the consultant can also say: 'A very nice house' even if he does not like the house."* However, some of the advisees see a contrast between discussing the property and their actual goal of taking an advisory service: *"At the end, it's about money. It was very nice that he showed me on the map that the house is in the countryside, but my concern is that I do not want to work until I'm 90 years old to pay it back."* The vast majority of the advisors finds the projection of the map a positive feature. An advisor argues that the visually supported discussion of the property *"introduces a new tension into the conversation such that it's not all that dry".* And another one claims that without referring to the house *"the empathy or so is not there, because it's just facts that you gather together".* However, some advisors are more critical: *"The map is a nice tool, a nice gimmick, but I do not think that it makes the customer feel more understood. Or that I am more interested in him. It is rather, so that one sees where the property is. That's a wow-effect."* The advisors and the advisees agree that the interaction around the map generates positive emotions, but the relevance remains questionable.

*4.2.3 Visualization and value manipulation.* We join A and C after A has explained the basic concepts concerning the Swiss mortgage system. They focus on the composition of a mortgage loan that meets the situation of C. This implies finding a balance between the more expensive but stable fixed mortgage interest rate and the more attractive but volatile LIBOR rates. An optimal solution yields a mortgage that remains affordable to C over a long period of time while keeping the interest rates as low as possible.

```
350      A and C look at the diagram projected on a paper
351  A:  then it would be °affordable° for you [.] then it would be ok °regarding° the affordability and
         feasability [0.4]
352  C:  mhm [0.3] o:kay [1.2] °its better [.] sounds good yes° [0.4] a:nd ah::m [1.8] oh no [0.4] °fine°
353  A:  please ask [0.4]
354      C looks down, focused on the projection, A looks at C
355  C:  no [.] its fine I figured it out [0.3] hahah [0.9]
356      C looks up to A and smiles, A looks at the C and smiles back, then bends over the table
357  A:  so [.] in case of those two we talk about [0.2] libor mortgage with one percent interest rate [0.6]
         and the one down here would be the:n a ten years fixed mortgage [0.3] four hundred thousand francs
         [0.3] with one point six percent [interest rates [°ha°
358  C:                                   [°mhm°        [yeah
359      A points to parts of the diagram showing two slices of a mortgage, C's gaze follows
360  C:  °and° the monthly rate would be a: then [0.4] °thousand two hundred twenty°
361      C points to a number at the bottom of the diagram, A looks first at C then at the diagram
362  A:  in the first [0.3] ten years i would say no:w [0.4] it is clear its a snapshot of current terms
363  C:  [0.2] °mh° [0.3]
364  A:  [A's monologue about LIBOR - omitted for space reasons]
365  A:  we have the possibility [.] we can simulate he:re the expected interest increase [0.6]
366      A looks at the diagram, touches the paper and positions the dial token next to it, C looks at A
         then on the table
367  A:  egh::::m [0.8] rather conservative it would be around zero point two percent or point three percent
         in a year [0.4]
368  C:  mh::m [0.7]
369  A:  °we can quickly try° to do it [5.3]
370      A repositions the dial token, then rotates it a few times forward and backward with her left hand,
         while rotating she points to specific changes in the diagram with her right hand, C looks at the
         A's pen and at the A's actions applied on the table, C keeps nodding
371  A:  you: see yeah [.] so: [.] with zero point three percent its interesting now [0.4] with zero point
         three interest increase [0.5] you would yet have [0.3] already [0.3] a change [0.4] ah: of the
         rates [.] you can see in the first year [0.3] it would be: relatively stable [2.4]
372      A points to the respective parts of the diagram, A looks alternately at C and the paper, C
         alternates the gaze between the paper and A, A points at specific years in the diagram and
         continues talking in a long turn
```

   The interaction between C and A is clearly organized around the visualization. A points to the diagram and looks at it, while talking (350). C clearly signalizes being intensively occupied with the content of the graphics (352), which A interprets and offers help straight away (353). Even though C refuses this offer (355), A continues with an extended explanation of what is visible in the projection (357). Thereafter, C signalizes understanding of the diagram by contributing an interpretation of what he sees (360). Subsequently, A offers a long explanation of the LIBOR system (364) and uses the value manipulation abilities of LivePaper to illustrate how the dependency on flexible rates may influence the mortgage affordability. A announces her intend verbally (365) and non-verbally (366) before she is ready to actually adapt the numbers (370), She interweaves the announcements with a general explanation of the nature of LIBOR interest rates. Thereafter, A explains what changes with the increase of the LIBOR (371) – she first points out that there is a clearly visible change and then lists the changes that happen to the monthly cost over the years (371). A is able to change the numbers while at the same time pointing to specific effects with her pen, and simultaneously explaining those effects (371). C follows A's pen and her actions, thus focusing on the causes and effects addressed verbally (370, 372). He remains calm but very attentive to what is happening in the interaction area. The nodding and visible following the advisor's physical action is a conventional way of showing one's acknowledgement of someone else's work (i.e., the extensive explanation). Goffman has also identified similar signals when people initiate interaction and greet each other [10,25]. Based on this (possibly expected) feedback from the client, the advisor engages even more intensely in the explanation; other excerpts have shown that if such feedback is missing, the advisor is likely to change the focus of her action: instead of manipulating the numbers and explaining the mechanisms, she would concentrate more on the client and would care more about improving the high-touch aspects of the encounter.

The advisees are overall very positive about the possibility to visualize changes and adaptations immediately. An advisee stresses that this possibility encouraged additional questions and suggestions: *"One just had the better overview, and you were able to track down how the values change, how they affect each other. I also dared more to ask how it would be now with 80% income, because I knew there, now he does not need to enter everything again."* Another one points out that it supports the negotiation or the problem-solving style of an advisory service: *"And then you can just move or adapt the different tranches and things, until you agree on what you want, you have a lot more room for discussion. Just what you really want to create."* However, some clients did not understand the graphical representation straight away and were strongly relying on the advisor's help with interpreting it: *"I did not understand the bars concerning the affordability and loan-to-value ratio – I did not find them so clearly at the beginning, so I was struggling: 'What is that?' I just saw 20 or 30 or 40 percent. But when he said what the bright and dark colors mean, that it shows my particular situation, then I got it. It made a click."* The advisors provide more mixed opinions about the visualizations and the opportunity to manipulate them. An advisor suggests, the amount of information may confound the client: *"Basically they are fixated on the current situation, on what is in the next two to three years. Sometimes LivePaper may confuse the client because of the overflow. Sometimes there is almost too much information. What happens in 10 years? (...) What if I reduce my job by 80% in 2 years? What happens if I have an accident? These are all hypotheses. The more you can illustrate, the more unsettled the client."* However, some advisors value the dial functionality a lot: *"The most meaningful thing for me was to use the dial to show the equity or purchase price increase. For me is a huge added value. You can show the customer if something needs to be adjusted. And you can immediately try this out".* Overall, the clients are consistently positive about the visualizations and the easy affordances for manipulating them collaboratively, whereas the advisors differ in this regard.

*4.2.3 Physical paper handling.* Throughout the advisory service, the participants can move various objects around the table. The same holds for paper: dedicated paper sheets with special marking and normal paper can be used alike – they can be put on the table, removed and then put back. The dedicated paper sheets are bound to the projected content: whenever they are positioned on the table, the last state of the linked projection can be seen. We join A and C towards the closure of the advisory service when A engages in summarizing the encounter by putting various marked sheets next to each other and refers to them.

```
385 A: so: for sure again [.] the mortgage mix we did [0.9] but also the: the whole compilation [0.8]
386    A picks up a piece of marked paper from a pile, positions it in the middle of the table, next to
       another marked sheet, A looks at the papers and at the C, C gazes at A
387 A: so that you see them too [3.2]
388    A moves the paper a bit, C and A look at the projections
389 A: of the [mortgages
390 C:        [mhm
391 A: so that you see how we perceive the loan-to-value ratio [0.3] how we assess the affordability [0.4]
       and how we made the mortgage mix [0.6]
392    A points to the specific areas of the projections on the two pieces of paper and looks up to the
       advisee sporadically, C looks at the projections, examines them carefully and speaks slowly
393 C: o:kay [.] yeah [0.3]
```

The advisor introduces the activity shift by positioning one of the marked sheets used earlier in the middle of the table (386). Piling the paper sheets in a specific order emerged as one of the practices, advisors employed to keep track of what appears on each piece of paper. She calls the attention of the advisee by verbally referring to *seeing* (387) and by moving the sheet a bit (388). In doing so, she leverages the material nature of paper to establish a situation where she and her client will look at the same piece of paper. Both participants gaze at the paper sheets in the middle of the table, while the advisor uses her pen to refer to particular content (392). The advisee takes some additional time to examine the content of the projections before he finally agrees (393). The interplay between verbal, non-verbal, and material signals to highlight the salience of an artefact in the interaction area has ritual aspects: it is not only commonly seen in religious ceremonies, but humans use it in plain interactional situations, e.g., when holding and moving a credit card or a purse on the restaurant table to catch the attention of the waiter.

The advisees refer to the handling of paper only rarely in the interviews. However, the opinions collected tend to be positive: *"I thought that was pretty good. I thought it was good that she could get a piece of paper* *back on the table and there was still the same information on it. And it was quite a natural way of dealing with the paper. It was nothing to pay extra attention to, for example to the stabilization that everything has to stop. I thought, it was very well solved.*" However, some advisees were confused by the name and the focus on paper itself: "*I would not call it LivePaper. I do not find the name so good. LivePaper makes you imagine an interactive paper, but here the paper has three circles on it and some strokes. I think the paper is not the lively main element. The greatest point is the light and shadow. The paper itself does not look so good.*" An advisor acknowledges the fact, that the linking between content and a paper sheet helped her structure the advisory service and the common interaction space: "*If the paper would only have served as a background, and I could not have taken the paper away, then there would be much less space. So I think it's great that I could put away the first sheet, and then just when I come back to it I can integrate it easily into my service, and it is then still has my inputs and that my drawings were still in the same place, that was great.*" Overall, the opinions suggest that paper handling was natural and easy to combine with ritual performances, but sometimes the clients would even expect more interactive features.

## 5 DISCUSSION

The results, in particular, the excerpt from the interaction, make clear that advisory services rely on physical and ritual interaction. We observe ritualized sequences of interaction: summarizing the encounter by referencing all previous steps, mutual assurances of the correctness of the notes taken, signalizing intents and activity shifts in a gentle way. Further elements of advisory service ritual involve the advisor's leading and structuring the encounter or assigning turns ("please ask"). When explaining, the advisor talks slowly and maintains the mutual gaze with the client; when talking about location, she provides additional information not explicit on the map. She maintains her face of a host, a representative of the bank, and an informed individual. The client engages in the rituals while taking the complementary position: he explains his situation while stressing elements, he assumes, might be relevant to the advisor (e.g., that he is <u>employed</u>), listens carefully to the advisor's monologues and tries to comprehend the information. The rituals and the attached roles get even more clear when one considers the physical conduct: the advisor is the one who distributes access to the interaction area, organizes the material in this area, and introduces new elements in an ordered fashion. Such practices as piling and keeping order in the pile show that the advisor finds it important to show herself as organized person; whenever she could expect problems, she implicitly or explicitly excuses it ("let's check") and signalizes her intentions even more through movement than through verbal cues (touching the dial token several times). What emerges is a choreography of performances that allow for transfer of information as well as emotional rapport building, the high-touch.

Importantly, the choreography emerges in a natural way, without explicit coordination effort. The data (both the data presented above, and the overall data considered for the analysis) do not include statements showing that the parties need to agree on how to structure their work or need to explicate this structure. There is no discussion or implicit negotiation of who is talking or doing what, when, and why – the interaction comes naturally to both partners, which supports the stance that the parties enact an interaction ritual [10,25]. The parties seem to share knowledge about what actions belong to the rituals they enact, even though most of the advisees had never been to a financial advisory service before. We argue, that the advisees could rely on the general knowledge of rituals in advisory services they might have acquired through other forms of consultations. This confirms the intuition that various instances of institutional talk do not only share rough characteristics, including the distribution of roles between a service representative and the advisee [30,59,71], but that they involve the same range of ritualized interactions. Practitioners often stress the fact that their domain or clientele are particular and they see little potential in transferring knowledge and ideas from other fields or view their specific goals and behaviors as distinct [15,67]. However, at the level of interaction rituals, a common ground seems to exist – the advisees are able to transfer and reuse behaviors they learned in other domains without requiring an explicit introduction and instruction for how to behave in financial advisory service. This bears potential for the generalization across different institutional encounters: from medical consultation to financial domain to employment agency.

CSCW research has long tradition in studying expert-layperson interaction in various domains [23,28,32,51]. In this regard, it often focused on identifying typical, domain-specific work practices and on understanding how IT interferes with those practices [14,16,22,51]. With pursuing the ritual perspective, this manuscript stresses the social and interactional aspects of institutional encounters: according to this framing, advisory services are about answering to social and interactional expectations or obligations rather than about completing a task or conducting work. While some articles already identified relationship building or impression management as key elements of advisory services [15,28,51], the design of systems for institutional settings was driven, primarily, by the task-oriented description [23,29,32,51]. LivePaper's design acknowledges the primacy of socio-material and physical interaction over task-related work; it affords rituals that respond to the generic social expectations and obligations. For instance, it affords an extended discussion about the property, such that the advisor can better fulfil the client's expectations of an empathetic and emotional interaction. It, also, affords the visualization and explanation sequences, such that the encounter responds to the client's expectation of meeting a knowledgeable expert, who clarifies complex issues for them. Showing empathy, positioning oneself as an expert, listening carefully and in an attentive manner are rituals common for high-touch interaction. They are ruled by social and general (rather than domain-specific) expectations and obligations. Comments collected from the advisees show, that what they valued a lot was not necessarily a successful completion of a mortgage service, but rather "clear and tidy" handwriting, "more room for discussion", or "emotionally involved" interaction. Whereas the CSCW focus on work practices has brought about insights and successful designs for expert-layperson encounters, we want to stress the role of social rituals and argue that designing for those rituals is the way to address the high-touch aspects of collaboration across settings and domains.

LivePaper does not only preserve rituals that existed before but also introduces new means to enhance the communication. The quantitative results show that advisees experience an improvement of the pragmatic quality as well as the qualities related to the high-touch character of the encounter (identity, stimulation). The comparison against the opinions from the interviews offers a clarification for enhanced pragmatic quality. The majority of advisees sees the overview and clarity as a great advantage of LivePaper advisory services, which are pragmatic qualities. They refer to an enhanced ability to understand things thanks to the visualizations and the value manipulation. In this context, they refer to qualities typical for experiential learning [43,44]: seeing what changes under what conditions [29]. LivePaper integrates this experience directly in the conversation about the client's situation rather than establishing a microworld [29]. Advisors who do not need to learn about these relationships anymore might lack the feeling of enhanced pragmatic quality. The collected data suggests that the existing usability problems might have compromised on the advisor's perception of the pragmatic quality: even though the tool produced appealing visualizations and allowed them to interact with the client in attractive ways, they also noticed that some graphics were overloaded with information and could confuse the client. Furthermore, some advisors experienced minor technical problems during the experiment, which could have impacted their overall assessment of pragmatic qualities. Both, advisors and advisees, on the other hand attribute improvements in emotional dimensions to the LivePaper. Thereby they refer to the multimodal experience of light, shadow, three-dimensional figures, etc. [4,45] which generates new sensory experiences. This confirms previous results on the use of multimodal tools in advisory services to enhance advisees' involvement [13] and have a lasting impact [12,23]. However, it remains open whether this improvement will hold on even if technologies comparable to LivePaper will spread in service encounters. While we cannot provide an answer yet. Because the collected data did not yield any differences between the first, second, and last treatment, we argue that the enhanced hedonic qualities result not simply from the novelty of the tool but from the way it is being used by the advisor and from the mixture of old practices and new technologies. In fact, the advisors and the advisees attest that multiple system features generate an experience of a *normal* advisory service: the advisors engage in writing, moving papers, introducing new elements and using pen as a pointer [15]. Even though the values on marked paper can be manipulated, which is different from conventional service, the values remain stable even when the paper was put away for some time according to object constancy [47,62,72], which is just the same conventionally. The collected opinions show that the advisees and the advisors consider such interactions essential for an advisory service.

LivePaper takes on a challenge addressed previously by numerous approaches: supporting advisory services without compromising on the interpersonal relation. To combine high-touch and high-tech, it proposes a paradigm based on preserving the rituals in physical terms and introducing singular elements that allow for digital augmentation. This paradigm could hold on to the challenge: the system shows how to combine enhanced transparency [53], better client education [29], and the feeling of joy [52] with paper-oriented material conduct typical for financial advisory services [15]. Whereas previous research reported on negative effects of digitalizing information during advisee's narratives at the beginning of the service [40–42], the support for note taking allows for a more fluent storytelling and offers means for signalizing full attention (e.g., closing the pen). Whereas earlier research reported on low advisees' involvement in various phases of the encounter [13,14], LivePaper uses crisp and dynamic visualizations to enhance involvement with the service content and symbolic means to enhance emotional engagement. Whereas previous research presents collaborative applications for advisory services as dangerous, ineffective and incompatible with the rituals [51,57], LivePaper makes clear that it is not a necessity. We argue, it is key to incorporate the physical side of rituals into the design. Thereby we mean not only supporting particular performances but also considering the physical configuration that goes beyond the interaction area [18]. All aspects starting from the environment of the interaction down to the particular movements or (as in LivePaper) shortcuts consisting of specific letters define build up rituals. And participants attend to advisory services with expectations and obligations in mind, which fit those rituals. For instance, one of the interviewees mentioned, he would expect the advisor to make more notes, but this expectation was not fulfilled to the last extent – it would be even less possible to fulfill it, if the advisor would have to write full words or to handle a complex system to enter the information. In fact, advisors also feel the obligation to give the client a feeling of being at the center of the encounter. Earlier reports make clear that advisors refuse to use systems like those if they cannot stand up to their obligations when using the system. This was observed in banks [40,67], in medical institutions [51], as well as in police [16]. LivePaper, even if not yet rolled out, has the best chance to be well adopted to the advisor's daily practice.

## 6 CONCLUSIONS

The insights do not come without limitations. As in other research-through-design undertakings [75] embedded in industry projects, the observed effects cannot be attributed to singular design decisions but rather to a bunch of decisions and assumptions implemented in the system. This compromises on internal validity: the constructs we identify and describe are likely to be interdependent and we stress the fact, that transferring them to other domains requires a holistic approach. The external validity, even though taken care of in form of realistic design experiment, can be enhanced as well through rolling out and conducting a pilot test of the system in advisory services with real clients. The analysis presented in the current manuscript focuses on the qualitative description of the interaction in LivePaper advisory services – conducting experiments with further test participants and coding for the effects identified here would further support the evidence and allow for direct comparison with conventional encounters. Since LivePaper was developed and tested with a single organization (even though advisors were coming from different branches), it may include bias towards specific work processes propagated in MoBa. Testing LivePaper with other banks and organizations could reduce this bias significantly. Finally, the current manuscript introduces the paradigm of designing for physical-ritual collaboration only in rough terms sufficient to motivate LivePaper's design. We intend to provide clear guidance for physical-ritual design approach in the future. LivePaper is a product of a three-year research program on improving various aspects of an advisory service and this particular study focuses on balancing the high-touch and high-tech aspects – we made our best to provide as much context information as possible to make this study easy to understand and follow. However, the full picture of LivePaper's abilities emerges when other articles are considered [15,17].

The current manuscript makes the following contributions. It offers a major improvement concerning the digital support of advisory services. Primarily, it shows that supporting service interaction with IT can be combined with enhancement regarding the high-touch character. So far, high-tech and high-touch were often considered contrary in the context for face-to-face services [2,74]. While CSCW has addressed institutional interaction in several contexts (doctor-patient, teacher-student, etc.) [24,29,51], it hardly ever went beyond analyzing the social interaction between the parties, such that only few IT-based solutions were presented [23,42], and yet more sporadically systems were tested that improved the interpersonal character of the institutional interaction [28]. Furthermore, this paper derives design requirements for a system to support co-located advisory services from previous literature and attributes them to the physical configuration. It also describes a system designed in accordance with those requirements. In a wider scope, the manuscript adds to the discourse on supporting expert-layperson collaboration including doctor-patient or teacher-student encounters: while recent studies blame the ineffective design and ignorance towards real-world practice for failure of the systems rolled-out to practitioners [16,40,51,57], this manuscript takes an alternative approach and suggests special care of social and physical rituals in digitally-supported encounters. In our opinion, the area of expert-layperson-technology interaction remains under-explored and needs now more attention than ever, because expert-layperson interactions will increasingly rely on AI and other IT instruments. Understanding what requires further attention when designing for expert-layperson encounters is more urgent than even before. The manuscript shows how the combination of social and material lenses contributes to better understanding of the problems in advisory services, thus aligning with practice-oriented CSCW research [9,15]. However, this study proposes a more advanced lens. It calls for a more careful study of generic rituals (in addition to the domain-specific practices): CSCW should attend more to the social expectations and obligations which affect interaction and offer designs that accommodate for those expectations. This approach is likely to produce more generic solutions and yield insights of general nature with potential to impact reference disciplines like sociology of interaction. This paper shows that it was the right approach to bridge the tension between high-touch and high-tech tendencies surrounding financial advisory services. Overall, the manuscript contributes knowledge of high practical relevance to specific application areas and describes a theoretical lens to be extended and tested through further research.

## ACKNOWLEDGMENTS

The authors thank all project members who have substantially contributed to turn the concept of LivePaper into reality. In particular, we thank Fiona Nüesch, Ulrike Schock, Marianne Wildi, André Renfer, and Mehmet Kilic for their intensive work and continuous support. We, also, thank all advisors and test clients for their participation in the experiments. This work was possible thanks to the financial contribution of the Swiss Innovation Agency (17716.1 PFES-ES) and support from the Hypothekarbank Lenzburg AG.